\documentclass[twocolumn,showpacs,preprintnumbers,amsmath,amssymb]{revtex4}
\usepackage{graphicx}
\usepackage{dcolumn}
\usepackage{bm}
\usepackage{multirow} 

\begin{document}

\title{\bf Tuning the bond order wave (BOW) phase of
half-filled extended Hubbard models\\}

\author{\bf Manoranjan Kumar$^1$, S. Ramasesha$^1$, Z.G. Soos$^2$}

\address{\it 
$^1$Solid State and Structural Chemistry Unit, Indian Institute of Science, Bangalore 5600 12, India,\\
$^2$Department of Chemistry, Princeton University, Princeton, N.J. 08544, USA}    
\date{\today}

\begin{abstract}
Theoretical and computational studies of the quantum phase diagram of the 
one-dimensional half-filled extended Hubbard model (EHM) indicate a 
narrow bond order wave (BOW) phase with finite magnetic gap $E_m$ for 
on-site repulsion $U < U^*$, the critical point, and nearest neighbor 
interaction $V_c \approx U/2$ near the 
boundary of the charge density wave (CDW) phase. Potentials with more 
extended interactions that retain the EHM symmetry are shown to have 
a less cooperative CDW transition with higher $U^*$ and wider BOW phase. 
Density matrix renormalization group (DMRG) is used to obtain $E_m$ 
directly as the singlet-triplet gap, with finite $E_m$ marking the 
BOW boundary $V_s(U)$. The BOW/CDW boundary $V_c(U)$ is obtained 
from exact finite-size calculations that are consistent with 
previous EHM determinations. The kinetic energy or bond order 
provides a convenient new estimate of $U^*$ based on a metallic point at $V_c(U)$
 for $U < U^*$. Tuning the BOW phase of half-filled Hubbard models with 
different intersite potentials indicates a ground state with large charge 
fluctuations and magnetic frustration. The possibility of physical realizations 
of a BOW phase is raised for Coulomb interactions.
\vskip .4 true cm
\noindent PACS 71.10.Fd, 71.30.+h, 71.45.Lr
 
\end{abstract}
\maketitle
\section{ Introduction}
\begin{figure}[h]
\includegraphics[scale=0.25,height=6.0cm,width=8.0cm,angle=-0]{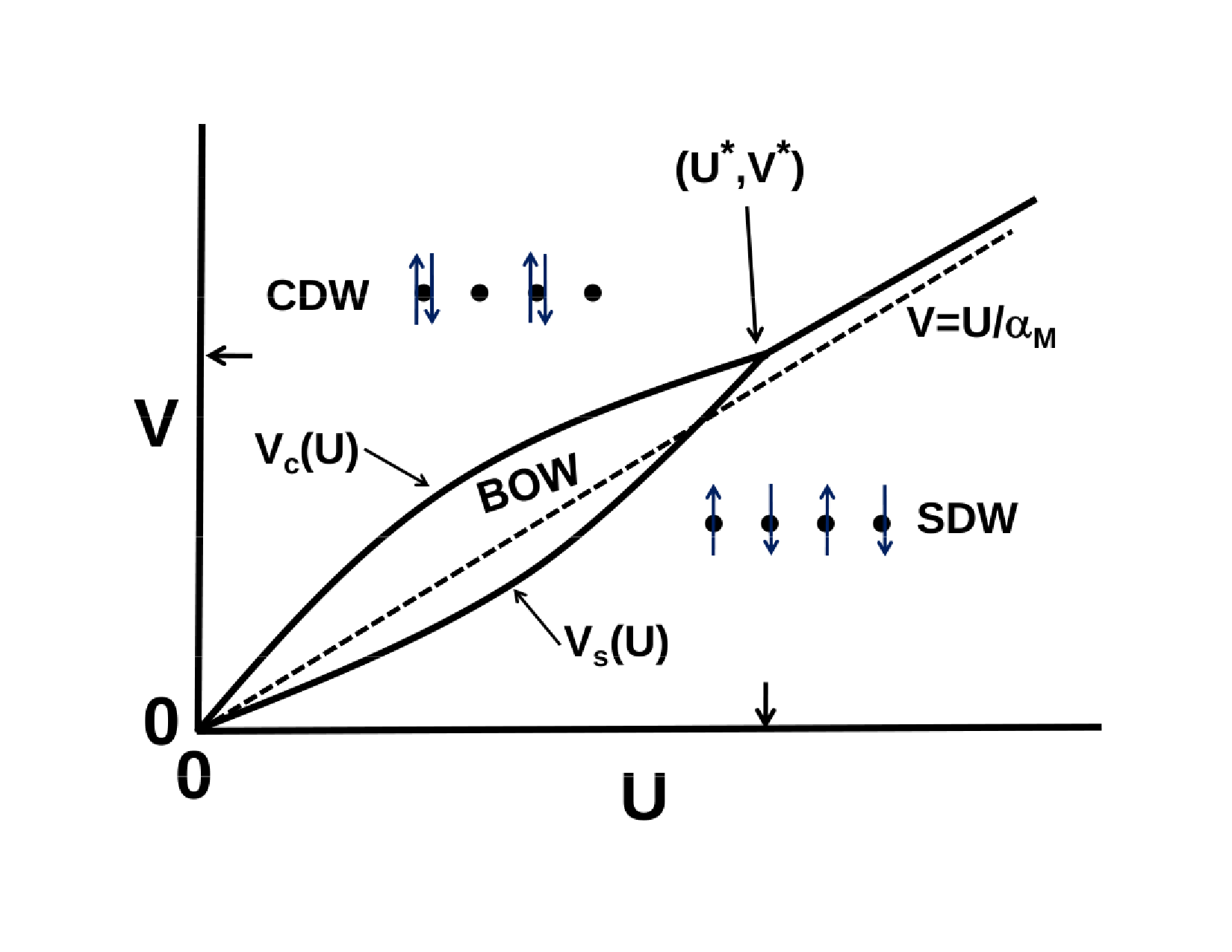}
\caption{Schematic quantum phase diagram of half-filled extended 
Hubbard models with on-site $U > 0$, nearest-neighbor $V$ and 1D potential with 
Madelung constant $\alpha_M$. The SDW/CDW transition line $V_c(U)$ is first 
order for $U > U^*$, the critical point, and second order for $U < U^*$. 
The BOW phase between $V_s(U)$ and $V_c(U)$ is a Mott insulator 
with a metallic point at $V_c$ and a finite magnetic 
gap, $E_m$, that opens at $V_s$. }
\label{fig1}
\end{figure}

 The quantum phase diagram of the half-filled extended Hubbard model (EHM) in 
one dimension (1D) illustrates competition among on-site repulsion $U > 0$,
 nearest-neighbor interaction $V$ and electron transfer $t$. Large $U$ gives 
a ground state (gs) with one electron per site while large $V > 0$ leads 
alternately to empty and doubly occupied sites. The gs phase diagram and 
its critical point $U^*$ have evolved since Hirsch's original study \cite{r1} of the 
boundary between a spin density wave (SDW) at large $U$ and a charge density 
wave (CDW) at large $V$. Nakamura \cite{r2} first proposed a bond order wave 
(BOW) phase between the SDW and CDW up to a critical $U = U^*$. The schematic quantum 
phase diagram in Fig.\ref{fig1} follows Sengupta {\it et al.} \cite{r3} 
and recent works by Zhang \cite{r4} and by Glocke {\it et al.} \cite{r5}. The CDW 
boundary that we denote as $V_c(U)$ has readily been found numerically. 
The value of $U^*$ and the SDW boundary at $V_s(U) < V_c$ are more challenging. 
The BOW phase is much narrower than sketched and lies between \cite{r4,r5} $V_s$ 
and $V_c$ for $U < U^* \approx 7t$. It is insulating and has a finite magnetic gap, 
$E_m$, that is expected to be very small because $V_s(U)$ is a Kosterlitz-Thouless 
transition \cite{r2}. Finite $ E_m$ implies {\it electronic} dimerization leading to 
long-range order at 0 K.

Multiple theoretical methods have been applied to the 
quantum phase diagram at 0 K. They include field-theoretical analysis of 
continuum models \cite{r2,r6}, weak-coupling expansions \cite{r7}, Monte Carlo 
simulations \cite{r1,r3}, density matrix normalization group (DMRG) methods \cite{r4,r8,r9} 
and the transfer matrix renormalization group (TMRG) \cite{r5}, 
in addition to exact diagonalization \cite{r10} of finite systems and 
perturbation expansions about the weak and strong coupling limits. 
The BOW phase for weak coupling (small $U,V$) has recently been demonstrated \cite{r9}
 for the EHM by functional renormalization group.

In this paper, we study the 0 K phase diagram of half-filled extended 1D 
Hubbard models with intersite interactions $V_m$ that are not restricted to $V_1\equiv V$. 
Any spin-independent $V_m$ retains \cite{r11} the translational, spin, and electron-hole 
symmetry of the EHM. Suitable potentials have electrostatic energy $ - V\alpha_M/2$
 per site, with Madelung constants $\alpha_M = 2$ for EHM, $2  \rm {ln2}$ for 
point charges, etc.
The CDW transition at large $V$ becomes less cooperative for 
$1< \alpha_M < 2$, thereby extending the BOW phase to larger critical $U^*$ and 
increasing the width $V_c(U)-V_s(U)$. We introduce a convenient new way 
to find $U^*$ and $V_c(U)$ based on the kinetic energy or bond order $p(U,V)$,
 which is maximized at a metallic point where the large $U$ and $V$ are balanced. 
The threshold of the BOW phase for $U < U^*$ is obtained by DMRG calculation 
of the magnetic gap $E_m$. Within the DMRG accuracy, finite $E_m$ at 
$V_s(U)$ below $ V_c(U)$ sets a lower bound on the onset of the BOW phase.

Competing interactions are widely invoked for materials whose gs depends sensitively 
on small changes \cite{r12}. The EHM literature 
deals with many-body techniques and ideas without regard 
to possible physical realization. Coulomb interaction between point charges is 
more physical. As in metal-insulator or Mott transitions, however, the assumption 
of purely { \it electronic} transitions is inherently an approximation. The SDW is 
unstable to a Peierls transition leading to dimerization for any harmonic 
lattice. Lattice dimerization corresponds to well-understood BOW phases 
that are realized in quasi-1D systems such as organic ion-radical 
salts \cite{r13} and charge-transfer (CT) salts \cite{r14} or conjugated 
polymers \cite{r15,r16}. Such systems have intermediate correlation $(U-V) > 2t$. 
The present study is restricted to a rigid 1D lattice with purely electronic 
instabilities. Possible physical realizations are mentioned in the Discussion.

 Previous EHM studies \cite{r1,r2,r3,r4,r5,r6,r7,r8,r9,r10} have focused 
on gs correlation functions for charge, spin or bond order as well as 
various susceptibilities. Our DMRG calculations target instead $E_m$ 
directly as the singlet-triplet gap, $E_{ST}$, between the lowest triplet state and singlet gs. 
We also present exact finite-size 
results that have been used to model the neutral-ionic transition of 
CT salts \cite{r17}. Contributions due to $t$ are particularly important 
at $V_c(U)$. The line $V_c(U)$ up to $U^*$ that marks a continuous 
transition is a metallic point with special properties according to 
the Berry-phase formulation of polarization \cite{r18,r19}. 
The motivations of the present work are to increase the range and 
width of the BOW phase by changing the intersite potential. In addition 
to the computational advantages of a wide BOW phase, control of the width 
is a step towards understanding a poorly characterized gs. On the SDW side, 
we refer to previous mappings \cite{r20,r21} of Hubbard models onto effective 
spin Hamiltonians with frustration \cite{r22}. Different perspectives on the 
0 K phase diagram of EHMs are consistent with other EHM results that also serve as checks for 
long range potentials.

The paper is organized as follows. Extended Hubbard models with spin-independent 
intersite potentials are defined in Section II along with bond orders, 
phase boundaries and the metallic point. DMRG calculations for the magnetic 
gap $E_m$ and exact finite-size results for $V_c(U)$ and $U^*$ are presented 
in Section III and compared to prior EHM results. The range and width of 
the BOW increase for less cooperative potentials. Section IV briefly discusses 
possible realizations of BOW systems and interpretations in terms of charge 
fluctuations and magnetic frustration.
\section{Phase boundaries of Hubbard models with extended interactions}
We consider a half-filled 1D Hubbard model with periodic boundary conditions 
(PBC), interaction potential $V_m$ between $m^{th}$ neighbors, on-site repulsion $U > 0$ and 
nearest-neighbor transfer $t$,
\begin{eqnarray}
\hat H= &-&t \sum_{p,\sigma}  (\hat a^{\dagger}_{p\sigma} \hat a_{p+1 \sigma}  + h.c. ) 
 + \sum_{p}\frac{U}{2} \hat n_p( \hat n_p ~-~ 1) \nonumber\\
 &+& \sum_{p} \sum_{m>0} V_m (\hat n_{p}-1) (\hat n_{p+m}-1)
\label{eq1}
\end{eqnarray}
The number operators $ \hat n_p$ have eigenvalues 
$n_p = 0$, 1 or 2. As written, the interaction energy 
is zero when $n_p = 1$ at all sites, at density $\rho = 1$.  
The Hellmann-Feynman theorem gives the weight or density of doubly 
occupied sites in the gs as
\begin{eqnarray}
\rho_2(U,V)=(2N)^{-1} \sum_p {\langle \hat n_p(\hat n_p-1) \rangle}
= \frac{1}{N}\frac{\partial E_0}{\partial U}
\label{eq2}
\end{eqnarray}
where $E_0$ is the exact gs energy. Stoichiometry relates the densities of 
cationic sites (holes) with $n_p = 0$ and anionic sites (electrons) with 
$n_p = 2$ as $\rho_0 =\rho_2 = (1-\rho)/2$. The choice of $V_m$ is open, 
subject to the constraint of spin-independent interactions with $V\equiv V_1$. 
We note that
\begin{eqnarray}
V_m(a)=V\frac{(a+1)}{(a+m)} 
\label{eq3}
\end{eqnarray}
corresponds to the EHM in the limit $a\rightarrow -1$, to a point charge model 
(PCM) at $a = 0$, and slower decrease for $a > 0$ that mimic molecular sites 
in a delocalized charge model (DCM). The electrostatic energy per two sites of 
the CDW with holes on one sublattice and electrons on the other is
\begin{eqnarray}
E_M(a)=2\sum_{m > 0} (-1)^m V_m(a)\equiv -V \alpha_M(a)
\label{eq4}
\end{eqnarray}
The Madelung constant $\alpha_M$ defined in Eq. \ref{eq4} is easily evaluated 
analytically for integer values of $a$.

In the strong-coupling or localized limit $(t = 0)$, the SDW/CDW boundary 
is simply $U = V\alpha_M$, the dashed line in Fig. \ref{fig1}. The gs for 
large $U$ has $n_p = 1$ at all sites and two-fold spin degeneracy at 
each site, while the gs at large $V$ is a doubly degenerate CDW. 
The curvature \cite{r23} of $V_m(a)$ ensures that the gs at $t =0$ is 
either the SDW or the CDW. In this limit, we have 
$dV/dU\rightarrow 1/\alpha_M$ and the density $\rho_2$ jumps from 0 to 1/2 at 
$V_c = U/\alpha_M$ on increasing $V$ at constant $U$. For finite $t$, the 
discontinuity of $\rho_2$ at $V_c(U)$ decreases and vanishes at $V^* = V_c(U^*)$ when 
the transition becomes continuous.

Translational symmetry leads to uniform bond order $p(U,V)$, proportional to 
the gs expectation value of the kinetic energy part of Eq. \ref{eq1},
\begin{eqnarray}
p(U,V)=\frac{1}{2N}\sum_{p\sigma}{\langle (\hat a^{\dagger}_{p\sigma} \hat a_{p+1\sigma}+h.c.)\rangle}
=-\frac{1}{2N}\frac{\partial E_0}{\partial t} 
\label{eq5}
\end{eqnarray}
$U = V =0$ is a metallic point with $p_0 = 2/\pi$ for a half-filled 1D 
band. The bond order of extended Hubbard models has a maximum at $V_c(U)$. 
A first-order transition for $U > U^*$ is indicated by discontinuous $p(U,V)$ 
at $V_c(U)$, while a second-order transition for $U < U^*$ has continuous 
$p(U,V)$ that develops a kink at $p(U^*,V^*)$.

The modern theory of polarization in solids grew out of the necessity of
 incorporating PBC for practical supercell calculations 
\cite{r18}. Subsequent generalizations of the theory and its relation to 
Berry phases have diverse applications, including how to distinguish between 
metals and insulators \cite{r19}. The theory is applicable to correlated models 
with neutral-ionic transitions \cite{r24} or to models in Eq. \ref{eq1} with any 
potential $V_m$. For EHMs, we require the exact 
gs of 1D supercells with N sites and PBC to compute the expectation 
value \cite{r24}
\begin{eqnarray}
Z_N(U,V)={\langle exp\big(\frac{2\pi i\hat M}{N} \big)\rangle}
\label{eq6}
\end{eqnarray}
$\hat M$ is the conventional dipole operator for unit charge and unit spacing in 1D
\begin{eqnarray}
\hat M=\sum_p {p(\hat n_p-1)}.
\label{eq7}
\end{eqnarray}
Extended models in Eq. \ref{eq1} have inversion symmetry at each site and 
at the center of each bond. The corresponding symmetries for finite $N$ 
are reflections through sites or through bonds. Either symmetry ensures 
that $Z_N$ is real in general for EHMs and reduces 
$Z_N$ to the twist operator \cite{r18} with $\cos(2\pi \hat M/N)$ instead of 
the exponential in Eq. \ref{eq6}. Quite generally, we have $Z_N \approx 1$ 
when the gs consists largely of sites with $n_p = 1$ in the SDW phase 
and $Z_N \approx -1$ when the gs is a CDW with $n_p = 2$ on one sublattice, 
$n_p = 0$ on the other. Hence the sign of $Z_N$ changes as $V$ 
increased at constant $U$. The point $Z_N(U,V_c) = 0$ for $U < U^*$ 
corresponds to a metal \cite{r19} that separates two insulating phases.
\section{Magnetic gap and metallic point}
 In this section we study the gs of $\hat H$ in Eq. \ref{eq1} for potentials $V_m(a)$ 
using density matrix renormalization group (DMRG) and exact finite-size 
calculations. Since $\hat H$ conserves total spin, $E_m$ is the singlet-triplet
 gap between the singlet gs and the lowest triplet state. We take $t = 1$
 as the unit of energy and find $E_m$ as the difference between the lowest 
energy states with total $M_S = 1$ and 0. DMRG with PBC is found 
to be very accurate when the ring is expanded from the middle by two 
sites at each step. Earlier DMRG introduced one site in the middle and 
another at the end of the block spins \cite{r27}. This had the disadvantage 
of adding a bond between a new site and the first site, whose operators have 
already been renormalized many times. The accurate DMRG-PBC procedure treats 
the ring as two chains that are joined at the ends.
\begin{figure}[h]
\includegraphics[scale=0.25,height=8.0cm,width=6.0cm,angle=-0]{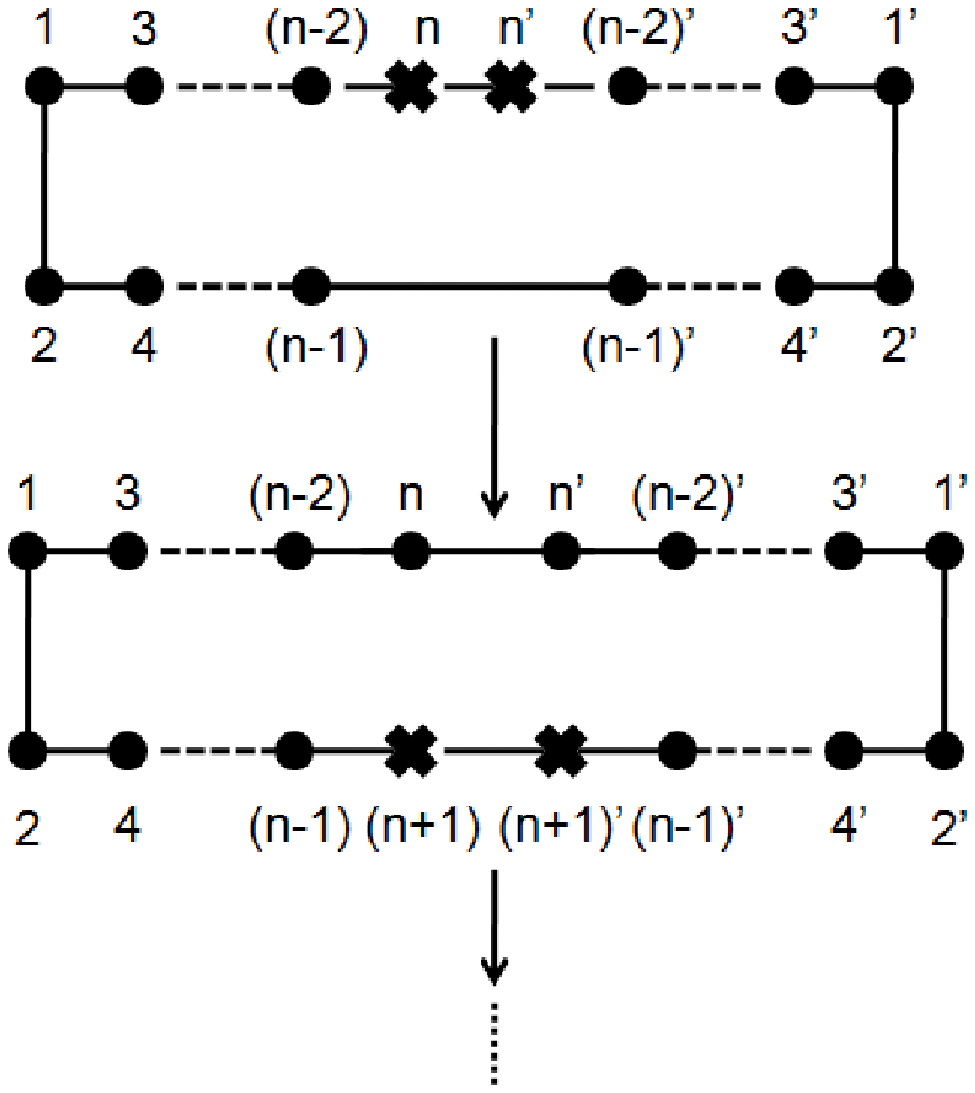}
\caption{Infinite DMRG procedure for a system with periodic boundary condition. 
Sites on left and right blocks are numbered as unprimed and primed integers 
respectively. Old sites are shown as filled circles and sites added at the DMRG steps are 
represented by crosses.}
\label{fig2p}
\end{figure}

 At each DMRG step, two sites are added, alternately, in the middle of one 
chain. The schematic diagrams are shown (see Fig. \ref{fig2p}). The new 
transfer terms at $2n$ system size are $(n,~~n'),~~[(n-1),~~ (n-1)'],~~ 
[(n,~~(n-2)']$ and $[(n-2),~~n']$. The electron-electron interactions are 
diagonal and hence are known to be accurate even when interacting sites are 
separated by several renormalization steps. The procedure introduces 
explicit transfers between sites whose operators have been renormalized 
only twice. We have carried out finite DMRG calculation for every $4n$ 
system size for increased accuracy. 

\begin{figure}[h]
\includegraphics[scale=0.25,height=8.0cm,width=6.0cm,angle=-90]{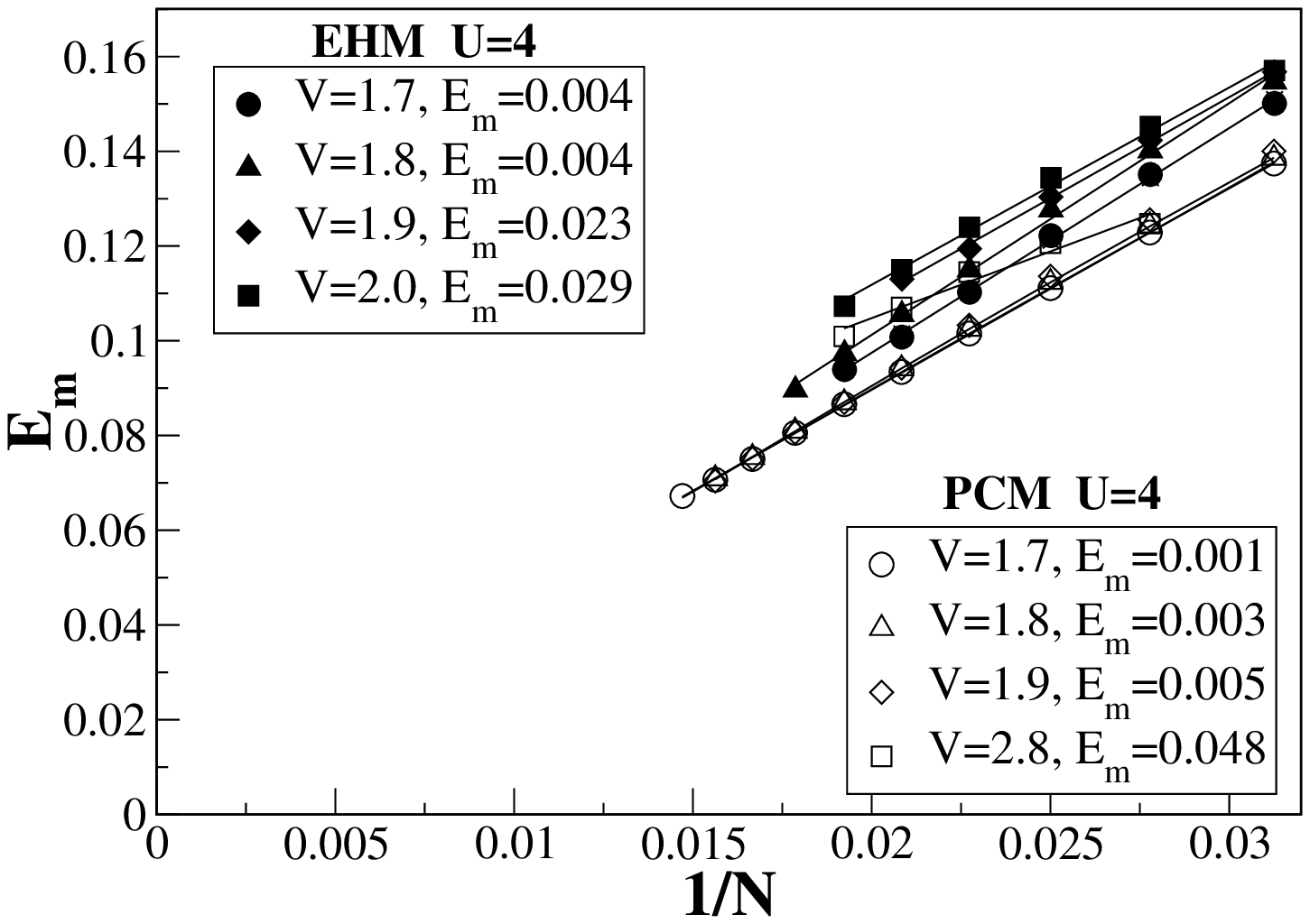}
\caption{ Singlet-triplet gap, $E_m$, of Eq. \ref{eq1} with $N$ sites 
and periodic boundary conditions vs. $1/N$ for EHM with $a=-1$ in 
Eq. \ref{eq3} and PCM with $a=0$. Extrapolation is between 
$N = 28$ and $N > 50$. The insets specify $U$, $V$ and 
the extrapolated $|E_m|<0.01$.}
\label{fig2}
\end{figure}

To calculate $E_m$, we retain $m = 150$ dominant density matrix eigenvectors 
for states in Eq. \ref{eq1} with either total $M_S = 0$ or 1. Each DMRG 
step gives $E_m(N)$ with $N$ increasing by two. We have performed DMRG 
calculations with finite ring size $N = 4n$ and find that one finite DMRG sweep 
is sufficient for good convergence of the energy. Fig. \ref{fig2} 
shows representative $E_m(N)$ vs. $1/N $ results up to $N > 50$ for the 
EHM $(a = –1)$ and PCM $(a = 0)$ at $U = 4$ for several values of $V$. 
Good $1/N$ dependencies are found and yield the extrapolated $E_m$ in 
Fig. \ref{fig2}. Other $U$ and potentials $V_m(a)$ show similar $1/N$ 
behavior for small $E_m$ when $V < V_c(U)$, while $V > V_c$ results 
in large $E_m \approx 1$ that increase rather than decrease with $N$. 
We estimate the accuracy of extrapolated $E_m$'s by comparison with 
$E_m = 0$ in the Hubbard model $(V = 0)$ or in the SDW phase $(V < V_s(U))$.
 The extrapolated $E_m/t$ are $\approx ±0.005$. Accordingly, we have 
assumed that the spin gap is finite when $E_m/t > 0.01$. 
We verified that $E_m$ does not depend significantly on the 
choice of $m$ for $m \ge 130$. Much larger $m$ leads to computational 
difficulties since the sparseness of the Hamiltonian matrix 
is significantly reduced. The $1/N$ dependence of $E_m(N)$ is 
monitored to find N after which the extrapolated value changed by 
less than $0.001$, and another iteration was then performed.

\begin{figure}[h]
\includegraphics[scale=0.25,height=8.0cm,width=6.0cm,angle=-90]{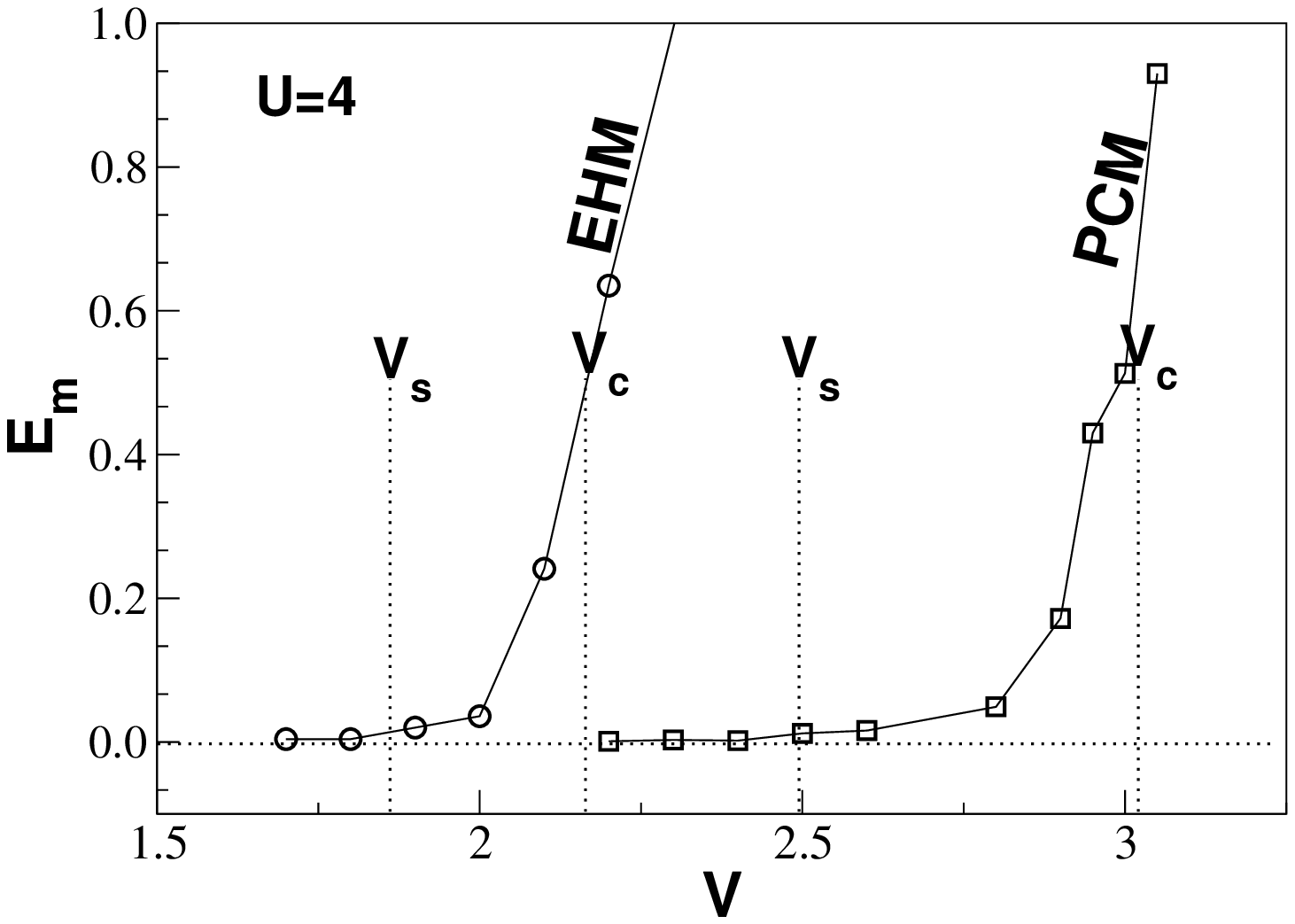}
\caption{Extrapolated singlet-triplet gaps, $E_m$, vs. $V$ for EHM and PCM at $U = 4$. 
The vertical dashed lines $V_s$ marks $E_m = 0.01$, the DMRG uncertainty (see text). 
The vertical dashed lines $V_c$ mark the metallic point $V_c(4)$ discussed in the text.}
\label{fig3}
\end{figure}

To show the threshold $V_s(U)$ for opening a magnetic gap, we plot $E_m$ 
in Fig. \ref{fig3} at $U = 4$ for EHM and PCM ($a = 0$ in Eq. \ref{eq3}). 
Dashed lines $V_s$ at $E_m = 0.01$ indicate the SDW/BOW boundary within 
our DMRG accuracy. The magnetic gap opens more slowly for PCM than for 
EHM and more slowly still for delocalized charges (DCM, $a = 1$, 
not shown). Dashed lines at $V_c$ indicate the BOW/CDW boundary that is 
found below. Similar DMRG evaluation of $E_m = 0.01$ for other $U$ and 
$V_m(a)$ combinations yield the $V_s(U)$ entries in Table 1. 
We find the EHM gap to open at $V_s = 1.86$ for $U = 4$. In this case, 
direct evaluation of $E_m$ is close to early estimates \cite{r2,r3} of 
the opening of the magnetic gap, while recent calculations \cite{r4,r5} 
give $V_s \approx 2.02$. The DMRG result in ref. 4 is based on a broad peak 
at $V_s$ that would presumably sharpen in larger systems. 
The TMRG result in ref. 5 is a thermodynamic method at finite temperature 
that nevertheless gives estimates for 0 K properties on extrapolation. We find a substantial 
 $E_m = 0.036$  for the EHM at $U = 4$ and $V = 2$. 

\begin{figure}[h]
\includegraphics[scale=0.25,height=8.0cm,width=6.0cm,angle=-90]{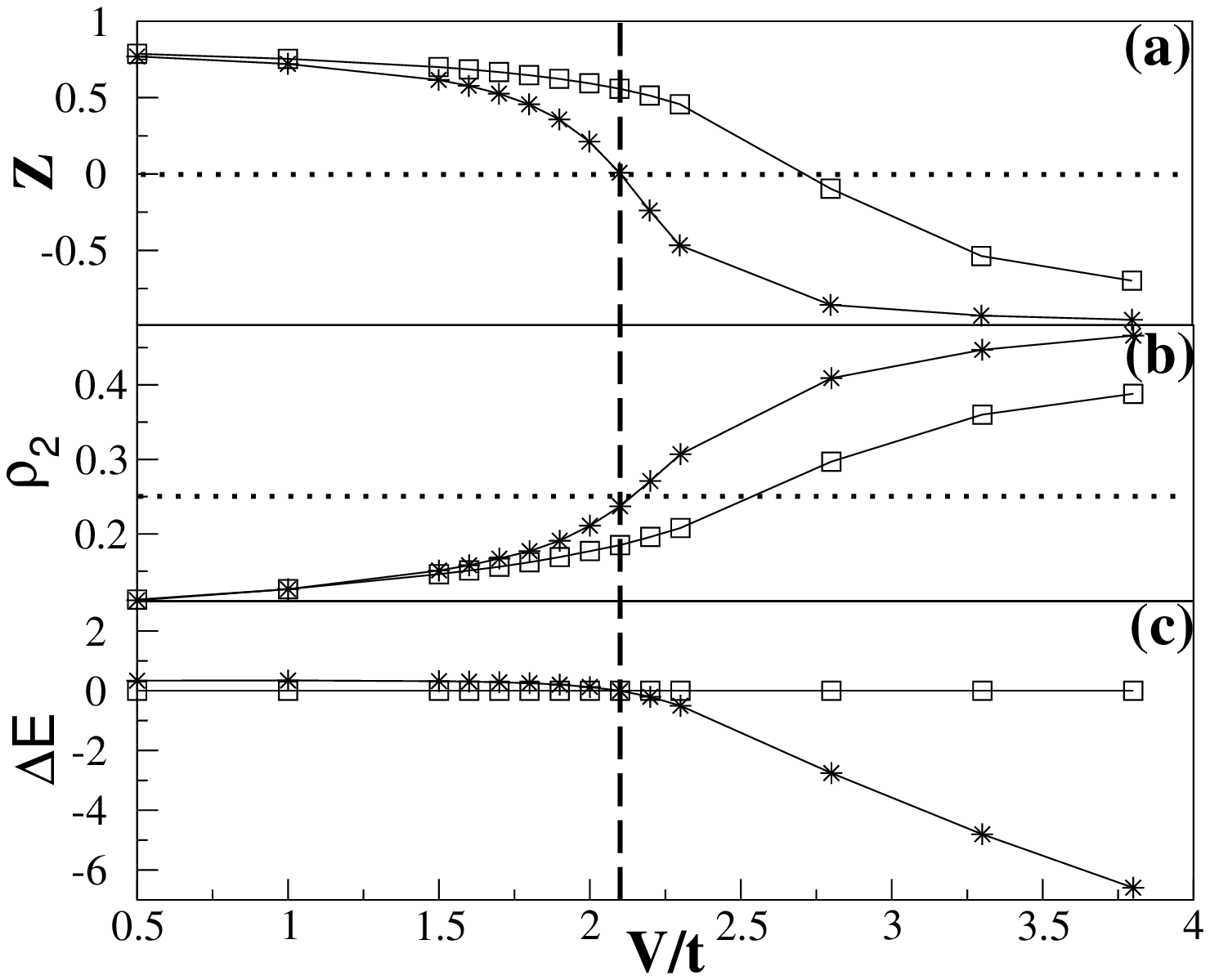}
\caption{Lowest energy singlets, both with e-h symmetry +1, for EHM with $U = 4$
 and 16 sites. Open squares, $B_u(k = \pi)$; stars, $A_g (k = 0)$. Panels 
(a) and (b) are $Z$ in Eq. 6 and $\rho_2$ in Eq. 2. Panel (c) shows 
the gs crossover, with $E = 0$ for $B_u$ and $\Delta E$ for $A_g$.}
\label{fig4}
\end{figure}

The CDW boundary at $V_c(U)$ has been found by diverse methods, 
as tabulated in ref. 8 for the EHM at $U = 4$; the range is 
$2.10 < V_c(4) < 2.16$. Here we evaluate $V_c(U)$ for finite 
systems of N sites and PBC in Eq. \ref{eq1}. Exact correlated 
states of Hubbard-type models, currently up to $N = 16$, 
are found using a many-electron basis of valence bond 
diagrams \cite{r28}.  As discussed in connection with the 
neutral-ionic transition in donor-acceptor stacks \cite{r17}, 
the gs has a symmetry crossover at $V_c(U,N)$ for $N = 4n$ 
that depends weakly $(\approx 1/N^2)$ on $N$. The bottom panel of 
Fig. \ref{fig4} shows the $U = 4$ crossover of the EHM 
from a singlet gs with even electron-hole symmetry that 
transforms as $k =\pi$ for $V < V_c(4,16) =2.10$ and 
as $k = 0$ for $V > V_c$. Moreover, we find 
$\rho_2(U,V) = 1/4$ and $Z_N(U,V) = 0$ for the $k = 0$ singlet 
at almost exactly $V = V_c(U,N)$ for $U < U^*$. The EHM crossovers 
at $U = 4$ for $N = 8$, 12, and 16 extrapolate to $V_c(4) = 2.16 \pm 0.01$,
 the entry in Table 1. Tighter extrapolation is possible 
using $N = 4n+2$ rings with antiperiodic boundary conditions \cite{r17}, 
but this was not pursued. Finite-size corrections to $V_c$ are even smaller 
for EHM at $U = 6$ or 10. They are somewhat larger for PCM, whose $V_c(U)$ 
entries in Table 1 have estimated uncertainties of $\pm0.02$, 
and larger still for DCM with $V_c(U)$ uncertainties of $\pm0.03$.

The boundaries of the BOW phase for several potentials are 
listed in Table 1. The SDW/BOW threshold $V_s(U)$ 
corresponds to $E_m = 0.01$. The BOW/CDW boundary $V_c(U)$ is 
the extrapolated gs crossover, the metallic point for $U < U^*$. 
The $U = 4$, 6 and 10 entries clearly show increasing $V_c-V_s$ 
as $\alpha_M$ decreases and the CDW transition becomes less 
cooperative. The demonstration of a BOW phase is then less 
demanding than for the EHM, where conflicting $U = 4$ results \cite{r3,r4,r5} 
were debated until recently. In particular, substantial $E_m/t > 0.2$ 
are achieved for the PCM and DCM at $V < V_c(4)$. Such gaps are 
robust numerically and easily exceed $k_BT$ in  
organic $\pi$-radical stacks \cite{r13} with $t\approx 0.1-0.3~eV$. 
It has been fully appreciated that an exponentially small $E_m$ poses numerical difficulties.
The present results are upper bounds for $V_s$ at finite $E_m$. The BOW phase
narrows at $U=6$ and is absent for the EHM at $U =10$. Our interpretation, continued in
the Discussion, is that a BOW gs requires large t in Eq. \ref{eq1} and strong charge
fluctuations. 
\begin{table}[h]
\begin{center}
\caption {BOW phase between $V_c-V_s$ for three potentials, 
EHM, PCM and DCM, with $a =-1$, 0, and 1 in Eq. \ref{eq3}, 
at $U = 4$, 6 and 10. The threshold $V_s(U)$ has magnetic gap 
$E_m = 0.01$, while $V_c(U)$ is the CDW boundary. 
The critical point $U^*$ has $V_s = V_c$ and no BOW 
phase. The Madelung constant, $\alpha_M$ in Eq. \ref{eq4}, 
decreases for less cooperative transitions}
\begin{tabular}{|c|cc|cc|cc|} \hline
\multicolumn{1}{|c|}{Models} & \multicolumn{2}{|c|}{EHM ($a=-1$)}
& \multicolumn{2}{|c|}{PCM $(a=0)$} & \multicolumn{2}{|c|}{DCM$(a=1)$} 
 \\
& $V_s$ & $V_c$ & $V_s$ & $V_c$ & $V_s$ & $V_c$ \\\hline
$U=4$   & 1.86 & 2.16 &2.50 &3.02 & 2.70 & 3.35\\\hline
$U=6$   & 3.06 & 3.10 &4.26 &4.45 & 4.63 & 5.03\\\hline
$U=10$  & 5.11 & 5.12 &7.33 &7.42 & 8.23 & 8.35\\\hline
\multicolumn{1}{|c|}{$U^*$} & \multicolumn{2}{|c|}{$6.7 \pm 0.2$} 
& \multicolumn{2}{|c|}{$10.6\pm 0.3$} & \multicolumn{2}{|c|}{$13.3 \pm 0.5$}
 \\\cline{1-7}
\multicolumn{1}{|c|}{$\alpha_M$ } & \multicolumn{2}{|c|}{$2$} 
& \multicolumn{2}{|c|}{$2 \rm {ln2}$} & \multicolumn{2}{|c|}{$4(1-\rm {ln2})$}\\\cline{1-7}
\end{tabular}
\end{center}
\label{table1}
\end{table}

We turn next to the critical point $U^*$, whose evaluation has 
been challenging. The most recent EHM values are $U^* = 6.7 \pm 0.2$ 
(ref. 5) or 7.2 (ref. 4), almost twice the first estimates from 
Monte Carlo simulations. We obtain $U^*$ from the kinetic 
energy at $V_c(U)$ as follows. The bond order $p(U,V)$ in 
Eq. \ref{eq5} is a continuous function of $V$ for $U < U^*$ 
with a maximum at the metallic point $V_c(U)$. The peak narrows 
with increasing $U$. A kink develops at $p(U^*,V^*)$ and $p(U,V)$ 
is discontinuous for $U> U^*$, where the transition is first order 
and there is neither a BOW phase nor a metallic point. Large $U$ 
generates a discontinuity in $p(U,V)$ at $V_c(U)$, with less kinetic 
energy on the CDW side. Second-order perturbation theory for the energy 
and Eq. \ref{eq5} provide a simple expression for the bond order,
\begin{eqnarray}
\label{eq8}
p^{(2)}(U,V)&=&\frac{4t\rm {ln2}} {V_c( \alpha_M -1)+(V_c-V)}, V< V_c  \\
p^{(2)}(U,V)&=&\frac{2t}{V_c(\alpha_M-1)+(V-V_c)(2 \alpha_M-1)}, V> V_c. \nonumber
\end{eqnarray}
Although $U \approx 10$ is not in the big-$U$ limit, Eq. \ref{eq8} 
rationalizes direct evaluation of $p(U,V)$ and the onset of a 
discontinuity at $U=V_c \alpha_M$.\\

\begin{figure}[h]
\includegraphics[scale=0.25,height=8.0cm,width=6.0cm,angle=-90]{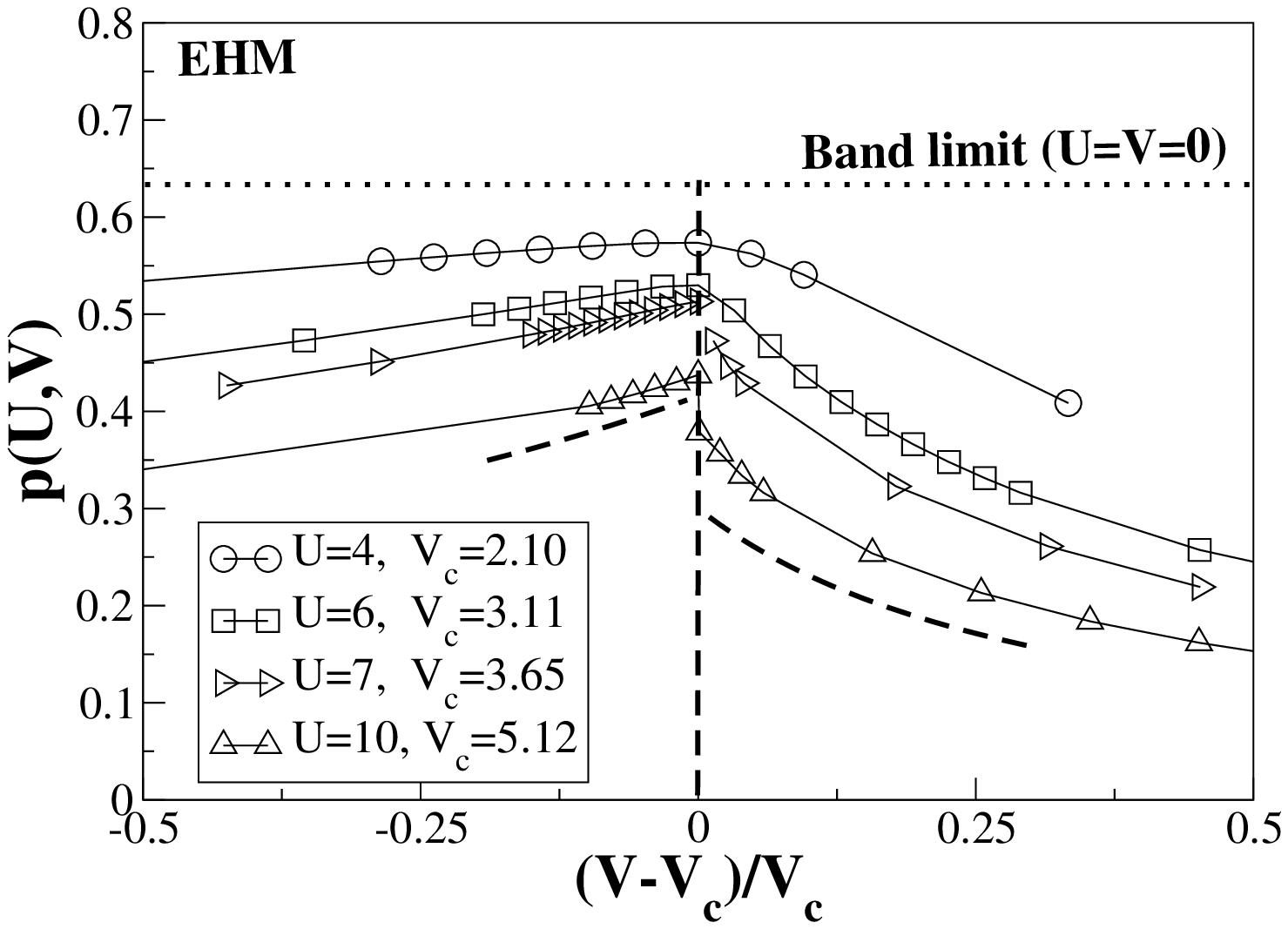}
\caption{Bond order, $p(U,V)$ in Eq. 5, vs. $(V-V_c)/V_c$ for EHM with 
$U = 4$, 6, 7 and 10. The dotted line is the band limit, $p_0 = 2/\pi$; 
the dashed line is the strong-coupling limit, Eq. \ref{eq8} with 
$V_c = 5.1$, $t = 1$, scaled by 0.80 }
\label{fig5}
\end{figure}

\begin{figure}[h]
\includegraphics[scale=0.25,height=8.0cm,width=6.0cm,angle=-90]{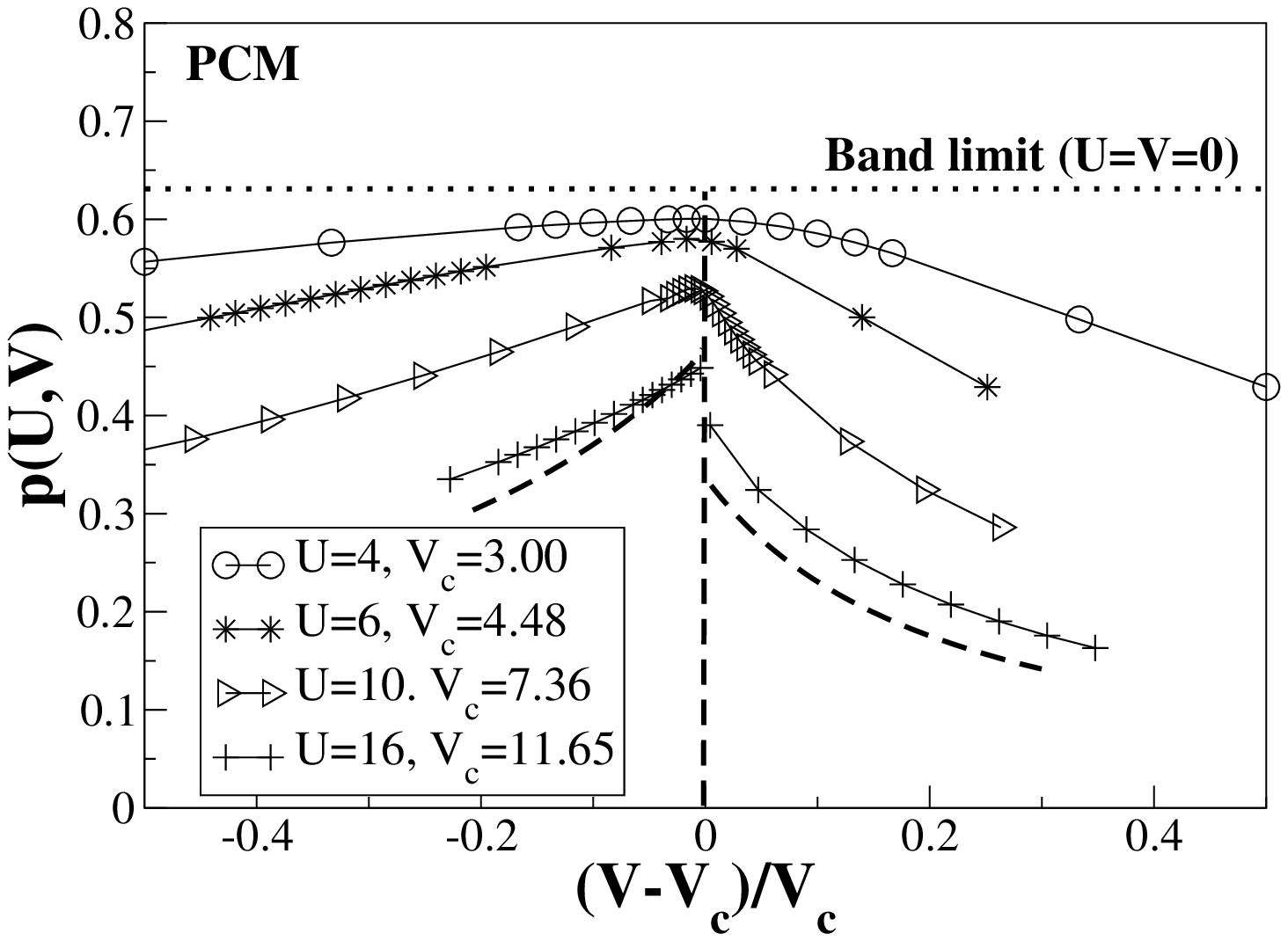}
\caption{Bond order $p(U,V)$ vs. $(V-V_c)/V_c$ for PCM with $U = 4$,
 6, 10 and 16. The dotted line is the band limit, $p_0 = 2/\pi$; the 
dashed line is the strong-coupling limit, Eq. \ref{eq8} with $V_c = 11.65$ 
and $t = 1$, scaled by 0.75}
\label{fig6}
\end{figure}

Figures \ref{fig5} and \ref{fig6} show $p(U,V)$ of $N = 16$ rings for 
the EHM and PCM, respectively, for the indicated values of $U$. The band limit 
($U = V = 0$) of the extended system is $p_0 = 2/\pi$; the analytical result 
for $N = 16$ is $1.3\%$ less
\begin{eqnarray}
p_0(16)=\frac{1+\sqrt{2}+2(\cos\phi+\sin\phi)}{8}=0.62842
\label{eq9}
\end{eqnarray}
where $\phi = 2\pi/16$. The band limit is the dashed horizontal 
line when plotted against ($V-V_c$)/$V_c$. 
In interacting models of Eq. 1,
 $p(U,V)$ has a broad maximum at $V_c$ that is less than $p_0$ and that 
sharpens with increasing $U$ for either potential. The kinetic energy at 
the metallic point decreases slowly. Careful examination of the $U = 6$ 
curve in Fig. \ref{fig6} shows that for $N = 16$, $p(6,V_c)$ is slightly 
larger for the gs on the CDW side. The $U = 4$ and 6 curves in Fig. 
\ref{fig5} show the same effect for the EHM. We used additional values of $U$ 
as well as $N = 12$ and 16 results to estimate $U^*$ as the onset of 
discontinuous $p(U,V)$. The $U^*$ entries in Table 1 increase as expected 
with decreasing $\alpha_M$. $U^* = 6.7$  for EHM agrees with the two recent calculations \cite{r4,r5}. The dashed line in Fig. \ref{fig5} for 
EHM is $p^{(2)}$ in Eq. \ref{eq8} with $V_c = 5.12$, scaled by a factor of 
0.80; the dashed line in Fig. \ref{fig6} for PCM is $p^{(2)}$  with $V_c = 11.65$ 
and a scale factor of 0.75. The perturbation result captures most of the $p(U,V)$ discontinuity.

The critical point $U^*, V^* = V_c(U^*)$ in Fig. \ref{fig1} marks the BOW phase with a metallic point. The 
first-order transition for $U > U^*$ is between two insulating phases. 
One has $\rho_2 < 1/4$ and $Z > 0$, the other has $\rho_2 > 1/4$ and 
$Z < 0$, and there is no metallic point. In the limit of large $U=\alpha_MV_c$ 
(or small $t$), we have $p \approx 0$ according to Eq. \ref{eq8}. The behavior 
of $p(U,V)$ and its weak size dependence provides a convenient new way 
to find $U^*$. The reason is that 
$p(U,V)$ varies slowly with V near the maximum at $V_c(U)$, 
whereas $\rho_2(U,V)$ or $Z(U,V)$ vary the most rapidly there. 
It is then numerically easier to discern a discontinuity in $p(U,V)$.

Half-filled Hubbard models with spin-independent potentials have e-h 
symmetry. Their dipole allowed (one photon) and two photon excitations 
are consequently to different manifolds of states, as discussed extensively 
for linear polyenes \cite{r30}. The lowest two-photon state, $2^1A_g$, 
has the gs symmetry and can be viewed as two triplets. Then $E_m = 0$ 
on the SDW side implies that the two photon gap also vanishes, while 
the one-photon excitation to $1^1B_u$ has finite energy. Conversely, finite 
$E_m$ in the BOW phase implies a finite two-photon gap. While DMRG for 
$E(2^1A_g)$ is less accurate at present than for $E_m$, we find a 
two-photon gap of 0.138 for EHM at $U = 4$, $V = 2.1$, well within the 
BOW phase. Moreover, since the gap is much smaller than $2E_m = 0.48$,
 we conclude that the triplets form a bound state. A finite two-photon 
gap is another signature of a BOW phase.
\section {Discussion}
We introduced potentials $V_m(a)$ in Eq. \ref{eq3} 
that retain the symmetry of the EHM in order to extend and 
widen the BOW phase sketched in Fig. \ref{fig1}. DMRG calculations 
of the magnetic gap $E_m$ are a direct new approach to the SDW/BOW 
boundary. The kinetic energy or bond order $p(U,V)$ at the metallic 
point turns out to be a convenient new way to evaluate the critical 
point $U^*$. Tuning the BOW phase by changing $V_m(a)$ also clarifies 
the electronic instability with increasing $U < U^*$ before the 
inevitable CDW transition at large $V$. The BOW phase is interesting 
but poorly characterized because it does not appear in the limit of 
either large $U$ or large $V$. It is an intermediate phase with strong 
charge fluctuations, competition between $U$ and $V$ leading to 
magnetic frustration, and fairly large $t$. Translational and 
inversion symmetry are broken in the BOW phase, while 
translational and electron-hole symmetry are broken in the CDW. 
Finite $V$ is required in either case.

The CDW shown in Fig. \ref{fig1} is a good representation of the 
gs when $V \gg U$ and sites are alternately empty and doubly occupied. 
The SDW in Fig. \ref{fig1} is a cartoon for $U \gg V$, however, since 
the gs of a Heisenberg antiferromagnet (HAF) is a linear combination 
of many states with $n_p = 1$ at all $p$. The Ne{\'e}l state 
shown has higher energy than the Kekul{\'e} functions, 
$| K1 \rangle$ or $|K2\rangle$, in which adjacent spins are 
singlet paired, either $2p$ with $2p-1$ or $2p$ with $2p + 1$ for all $p$. 
$| K1 \rangle$ or $|K2\rangle$ is a linear combination of $N$-spin states, 
half up and half down, that includes the Ne{\'e}l state. The actual gs 
for $U \gg V$ also has singlet pairing of sites that are not adjacent. 

 We have deliberately omitted a BOW cartoon in Fig. \ref{fig1}; it is usually shown
 \cite{r2,r3,r4,r5} as ($\uparrow ~~~\downarrow$)($\uparrow ~~~\downarrow$) to suggest
broken symmetry for bond orders. Such a representation, suitable for
a dimerized HAF, is quite misleading for a BOW gs with almost $50\%$ of
empty and doubly occupied sites (Fig. \ref{fig4}) and large bond order
 (Figs. \ref{fig5} and \ref{fig6} ) for V slightly less than
$V_c(U)$. Since t connects adjacent sites, large $p(U,V)$ indicates
strong mixing of a singlet pair at sites $p$, $p+1$ and ionic singlets
with $n_p=2$, $n_{p+1}=0$ or $n_p=0$, $n_{p+1}=2$. The BOW gs is a linear
combination of the full Hubbard basis of $4^N$ states, aside from particle
number or symmetry constraints. A product of dimer function (1,2)(3,4)... gives 
some insight into the BOW gs. The dimer function (1,2), an 
approximation for the BOW gs, is   
\begin{eqnarray}
(1,2) &=&[\frac{\cos \theta}{\sqrt{2}}(a^\dagger_{1\alpha}a^\dagger_{2\theta}-a^\dagger_{1\theta}a^\dagger_{2\alpha})\nonumber\\
&& +\frac{\sin \theta}{\sqrt{2}}(a^\dagger_{1\alpha}a^\dagger_{1\theta}+a^\dagger_{2\alpha}a^\dagger_{2\theta})]|0\rangle
\label{eq18}
\end{eqnarray}
where $a^\dagger_{p\sigma}$ is a creation operator in Eq. \ref{eq1} and $\theta$ is a variational parameter.
 The state in the first term is the covalent (Heitler-London) singlet; that in the second is the ionic singlet. The bond 
order per site is $\sin\theta \cos\theta \approx1/2$ for equal admixture at $\theta = \pi/4$. 
Such $p(U,V)$ are found directly for $U < U^*$. The product function indicates a BOW with large charge fluctuations

Strong coupling gives a first-order SDW/CDW transition at $V_c 
= U/\alpha_M$ for $t = 0$ and no BOW phase for $U > U^*$. Corrections 
for finite $t$ can readily be found. At the boundary, the energy for 
an electron transfer in either the SDW or CDW gs is
\begin{eqnarray}
\Delta E=U-V_c=U(1-\frac{1}{\alpha_M(a)})
\label{eq10}
\end{eqnarray}
We have $\Delta E > 0$ for $\alpha_M > 1$ and can apply 
perturbation theory in $t/\Delta E$. Second-order $(t/U)^2$ 
corrections to $E_0(U,V)/N$ on either the SDW or CDW side 
yield an approximate $\rho^{(2)}_2(U,V)$ according to 
Eq. \ref{eq2}. The jump at $V_c$ decreases and vanishes at
\begin{eqnarray}
U^{(2)}=\frac{2 \alpha_M t \sqrt{2\rm {\rm ln2}+1}}{ ( \alpha_M-1) }
\label{eq11}
\end{eqnarray}
$U^{(2)}$ is a simple estimate for $U^*$ that shows how potentials 
with small $\alpha_M > 1$ extend the range of continuous 
$\rho_2(U,V)$. We obtain $U^{(2)}/t = 6.18$ for the EHM with 
$\alpha_M = 2$. Second order corrections yield a comparably 
accurate estimate for a continuous neutral-ionic transition 
in CT salts \cite{r30p}. The values of $U^{(2)}$ are 11.1 for 
PCM and 16.7 for  DCM. Second order corrections yield 
$V_c(U,a) > U/\alpha_M$ but overestimate the increase; 
the fourth order correction \cite{r21} for EHM reduces the increase.

 Half-filled Hubbard models have spin-charge separation when $U - V$ 
is large compared to t. The gs on the SDW side can be mapped into 
an effective spin Hamiltonian for any potential $V_m$. Up to $t^4$,
 $H_{eff}$ is a HAF with $J_1$, $J_2 > 0$ and frustration; a magnetic gap 
opens \cite{r22} at $J_2/J_1 = 0.2411$. The possible connection between a 
BOW phase and spin frustration has been pointed out previously \cite{r21}, 
and it is a delicate matter. Seitz and Klein \cite{r20} obtained $H_{eff}$ up to $(t/U)^6$ 
for $V = 0$ in Eq. \ref{eq1} by systematically considering virtual 
transfers in the covalent sector with $n_p = 1$ at all $p$; an even number 
of transfers is required to end up with $n_p = 1$ at all $p$. 
Van Dongen \cite{r21} obtained $H_{eff}$ for EHM up to $[t/(U-V)]^4$. 
The $t^2$ term gives a HAF with nearest-neighbor exchange $J_1 = 4t^2/(U-V)$, 
and $J_1$ has contributions from all higher orders. The $t^4$ order 
introduces a second-neighbor exchange $J_2$, also AF, and all higher orders 
contribute to $J_2$. Mapping into a HAF with $J_1$ and $J_2$ breaks down 
in the next order, even if such corrections are included in $J_1$ and $J_2$, 
because $t^6$ generates exchange interactions among four successive 
spins \cite{r20}. At the $t^4$ level, the ratio $J_2/J_1$ can readily be 
generalized to the potentials $V_m(a)$ in Eq. \ref{eq3}. We obtain
\begin{eqnarray}
\frac{t^2F(a,V)}{(U-V)^2}=\frac{J_2/J_1}{1+4J_2/J_1}
\label{eq12}
\end{eqnarray}
\begin{eqnarray}
F(V,a)-1\equiv \frac{V-V_2(a)}{U-V_2(a)}=\frac{V}{V+(a+2)(U-V)}
\label{eq13}
\end{eqnarray}
The factor $F(V,a)$ reduces to $F = 1$ at $V = 0$ and to $F(V,–1) = 1 + V/U$ 
for the EHM. Finite $E_m$ requires $J_2/J_1 > 0.2411$ in spin chains when the 
rhs of Eq. \ref{eq12} is 0.1227. 

 Since the 1D Hubbard model is rigorously known \cite{r31} to have $E_m = 0$ 
for $U > 0$, $F = 1$ is insufficient to open a gap at $U/t = 2.85$ for $V = 0$ 
in Eq. \ref{eq12}. $H_{eff}$ with $J_1$ and $J_2$ clearly fails for $(U-V) < 3t$ 
because it incorrectly predicts finite $E_m$. 
The EHM has $F = 1.5$ at $V = U/2$, when a magnetic gap opens at $(U-V)/t = 3.5$ 
according to Eq. \ref{eq12} and yields $U^* = 7.0$ in surprisingly good agreement 
with recent calculations \cite{r4,r5} and Table 1. 
But the correction to $J_2/J_1$ goes as $[t/(U-V)]^2 \approx 8\%$ 
in the next order and $H_{eff}$ has an additional term. Potentials with $a > –1$ 
lead to $F > 1.5$ at $V =U/\alpha_M$ and to slightly larger $(U-V)/t 
\approx 3.6-3.7$ in Eq. \ref{eq12} that, however, substantially 
overestimates $U^*$ in Table 1. Hence we regard the EHM result 
to be accidental. Growing magnetic frustration with increasing 
V gives insight into how a BOW phase might develop, but the 
actual BOW gs for $U < U^*$ and $V < V_c(U)$ has finite 
$\rho_2$ that cannot be represented by a HAF with $J_1$, $J_2$. 

We consider next the possibility of a physical realization of a BOW phase. 
Several obstacles were mentioned in the Introduction. The material must 
have quasi-1D electronic properties that can be approximated by an extended 
Hubbard model such as Eq. \ref{eq1}; it must be close to a CDW instability; 
and yet it must avoid the Peierls instability of half-filled bands. 
Electrostatic (Madelung) energies are inherently 3D. The generalization of 
$\alpha_M$ in Eq. \ref{eq4} to 3D is straightforward and has been applied to 
CT salts \cite{r32} with neutral or ionic gs. The strong $\pi$-acceptor 
A = TCNQ (tetracyanoquinodimethane) forms an extensive series of salts 
that contain face-to-face stack of $A^-$ ion radicals \cite{r13}. 
1:1 salts correspond to a half-filled band, and 1D Hubbard-type models 
describe the magnetic, optical and electronic properties of TCNQ salts. 
Formally, Eq. \ref{eq1} refers to electrons in the LUMO of TCNQ with 
$t \approx 0.1-0.3$ eV. 

We recently pointed out that the $A^-A^-$ stacks in $M^+\rm {TCNQ}^-$ salts with
 $M$ = Na, K, Rb and Cs are close to the CDW boundary of $AA^{-2}$ stacks, 
thereby satisfying one condition for a possible BOW phase \cite{r33}. 
Alkali-TCNQ salts are insulators with Coulomb interactions among localized 
charges on $M^+$ and delocalized charges on $A^-$. Delocalization suggests 
that the appropriate $V_m(a)$ has $a > 0$ rather than $a = 0$ for point 
charges, although 3D electrostatic interactions have to be taken into 
account. In any case, we have {\it molecular} sites in Eq. \ref{eq1} and 
the approximation of a rigid lattice with uniform spacing must extend 
to rigid molecules. Alkali-TCNQ salts \cite{r34} have a phase transition
 around $T_d \approx$ 300 K to a dimerized stack whose gs is a BOW expected 
in systems with triplet spin excitons \cite{r13}. The $A^-$ stacks are regular for $T > T_d$ 
and have inversion symmetry at each $A^-$. A candidate BOW phase then refers 
to $T > T_d$ and must have $E_m \approx k_BT_d$ in order to justify using the gs 
at 0 K. The molar spin susceptibility, $\chi_{M}(T)$, of alkali-TCNQs is small 
\cite{r34} ($\approx 10^{-4}$ emu) at $T_d$, consistent with typical Hubbard-model 
parameters \cite{r33}. The behavior of $\chi_M(T)$ for $T > T_d$ in 
these salts, however, is not consistent with a HAF, the early model 
of choice, or with any Hubbard model with $E_m = 0$ and $\chi_M(0) > 0$.
 Finite $E_m$ in a stack with equal spacing accounts qualitatively 
for $\chi_M(T)$ at $T > T_d$. But magnetic data is merely suggestive of 
a BOW phase.

Quasi-1D materials based on stacks of planar molecules have strong 
electron-vibrational coupling to molecular vibrations as well as 
to lattice modes. The BOW phase is an electronic instability that 
breaks inversion symmetry at sites. Lifting inversion symmetry has 
spectacular vibrational consequences that have motivated Raman and 
infrared studies of dimerization transitions in many systems \cite{r35p}, including 
TCNQ salts. The appearance of a totally symmetric TCNQ-vibration, polarized 
along the stack, in the ir spectrum at $T > T_d$ is evidence for a BOW 
phase. Such a mode was reported \cite{r35} for a CN stretch in K-TCNQ and 
rationalized in terms of dimerization fluctuations before the suggestion 
\cite{r2} of a BOW phase for the EHM. Several coupled modes were reported \cite{r36} 
in powder ir spectra at $T > T_d$ of several alkali-TCNQs. The second form of 
the rubidium salt, Rb-TCNQ(II), is a good candidate for a planned single 
crystal ir study. It has a low \cite{r34} $T_d \approx$ 230 K and a crystal structure
 \cite{r36} with one molecule per unit cell at 300 K. 

In summary, we have characterized the BOW phase of half-filled extended Hubbard 
models with 1D potentials $V_m(a)$ in Eq. \ref{eq3}. The BOW phase (Table 1) 
for $U < U^*$ extends from $V_s(U)$ where the magnetic gap opens to the 
metallic point $V_c(U)$ at the CDW boundary. The kinetic energy or gs bond 
order provides an accurate new estimate of the critical $U^*$ that 
terminates the BOW phase. The SDW/CDW transition for $U > U^*$ is a 
first-order transition between two insulating phases. We find that the 
BOW gs has strong charge fluctuations, magnetic frustration and a finite 
two-photon gap. The present analysis of competing interactions is limited 
to electronic correlations in a 1D model. Both electron-phonon coupling 
and electron-molecular-vibration coupling will have to be considered 
to establish that several suggestive observations on alkali-TCNQ 
salts at $T > T_d$ can be interpreted as physical realizations of BOW systems.

{\bf Acknowledgments.} ZGS thanks A. Painelli for stimulating discussions 
about the BOW phase of the EHM. SR thanks DST India for JC Bose National 
Fellowship and a grant through project no. SR/S2/CMP-24/2003. MK thanks UGC 
India for a senior research fellowship.

\end{document}